\documentclass[twocolumn,preprintnumbers,amsmath,amssymb]{revtex4}

\usepackage{graphicx}
\usepackage{dcolumn}
\usepackage{bm}
\usepackage{eqname}  
\usepackage{color}
\usepackage{natbib}

\begin{document}


\title{Signatures of Fractional Quantum Hall States in Topological Insulators}
\author{Dong-Xia Qu$^1$\footnote{All correspondence should be addressed to D. Qu (qu2@llnl.gov)}}
\author{Y. S. Hor$^2$}
\author{R. J. Cava$^3$}
\affiliation{$^1$Lawrence Livermore National Laboratory, Livermore, CA 94550, USA\\
$^2$Department of Physics, Missouri University of Science and
Technology, Rolla, MO 65409, USA \\
$^3$ Department of Chemistry,  Princeton University, New Jersey 08544, USA.}
\date{\today}

%
%
%
%
%
%

\begin{abstract}

The fractional quantum Hall (FQH) state is a topological state of
matter resulting from the many-body effect of interacting
electrons and is of vast interest in fundamental physics
\cite{Tsui1982, Jain2000}. The experimental observation of
topological surface states (SSs) in three-dimensional bulk solids
has allowed the study of a correlated chiral Dirac fermion system,
which can host a single Dirac valley without spin degeneracy
\cite{Hasan2010,Hsieh2008,Hsieh2009,Roushan2009,Chen2009,Qu2010,Analytis2010,Xia2009}.
Recent theoretical studies suggest that the fractional quantum
Hall effect (FQHE) might be observable in topological insulators
\cite{DaSilva2011,Moore2010}. However, due to the dominant bulk
conduction it is difficult to probe the strong correlation effect
in topological insulators from resistivity measurements
\cite{Qu2010,Analytis2010}. Here we report the discovery of FQH
states in Bi$_2$Te$_3$ from thermopower measurements. The surface
thermopower is ten times greater than that of bulk, which makes
possible the observation of fractional-filled Landau levels in
SSs. Thermopower hence provides a powerful tool to investigate
correlated Dirac fermions in topological insulators. Our
observations demonstrate that Dirac topological SSs exhibit
strongly correlated phases in a high magnetic field, and would
enable studies of a variety of exotic fractional quantum Hall
physics and other correlated phenomena in this newly discovered
chiral Dirac system.
\end{abstract}


%

\maketitle

\newpage

Topological insulators (TIs) are a new class of quantum states of
matter with topologically protected conducting SSs, arising from
the topology of the bulk electronic band structure
\cite{Kane2005,Bernevig2006,Moore2007,Fu2007,Qi2008}. There are
two distinguishing features of topological surface states. One is
the existence of an odd number of Dirac cones on each surface, and
the other is the helical spin arrangement
\cite{Hsieh2009,Roushan2009,Xia2009}. Theoretically, the
relativistic nature of Dirac fermions is believed to significantly
modify the electron-electron interactions, with the possibility to
produce more robust ground states at the $n=1$ Landau level (LL)
in TIs than in conventional two-dimensional electron systems
\cite{Goerbig2006,Yang2006,Apalkov2006,Toke2006}. The unique spin
texture and the coexistence of non-insulating bulk states also
raise the intriguing question of whether TIs may host exotic FQH
states owing to the non-trivial Berry's phase \cite{Fu2007}, huge
Zeeman energy \cite{Liu2010}, and the screening effect from bulk
carriers \cite{DaSilva2011}. The potential realization of more
stable non-Abelian FQH states in TIs is of practical interest for
topological quantum computing \cite{Moore2010,Nayak2008}.

There have been magnetoresistance measurements at $n=0$ and $1$
LLs in (Bi$_{1-x}$Sb${_x}$)$_2$Se$_3$ \cite{Analytis2010} and at
$n=4$ and 5 LLs in Bi$_{2}$Se${_2}$Te \cite{Xiong2011}. However,
the sub-integer oscillations at $n=1$ LL in
(Bi$_{1-x}$Sb${_x}$)$_2$Se$_3$ can only be resolved in the
second-derivative trace and their best linear fit intersects the
filling-factor axis at $0$ instead of $1/2$, inconsistent with a
Dirac spectrum. Such a property makes the exact origin of these
oscillations unclear \cite{Analytis2010}. On the other hand, the
FQHE in TIs is theoretically precluded in the $|n|>1$ LLs
\cite{DaSilva2011}, suggesting the features in Bi$_{2}$Se${_2}$Te
unlikely to be ascribed to the FQHE. So far, transport studies of
FQHE in TIs have been limited to samples with mobility below 3,000
cm$^2/$Vs and the SS conduction is susceptible to conducting bulk
states.


To explore the existence of FQH states in TIs, we present
thermoelectric measurements on the Bi$_2$Te$_3$ crystals. The
surface mobility of these crystals is up to $14,000$ cm$^2/$Vs
\cite{Qu2010}, comparable to the Hall mobility ($30,000$
cm$^2/$Vs) of high quality graphene where the FQHE has been
recently discovered \cite{Dean2011}. We first examined the
dependence of thermopower $S_{xx}$ on temperature $T$ in both
metallic and nonmetallic samples (Fig. 1, insets). Though $S_{xx}$
shows a low-$T$ peak in all these samples, the peak of the
nonmetallic samples Q1 and Q2 is significantly stronger than that
of the metallic sample M1. These observed peaks indicate the
occurrence of phonon-drag effect that is expected to appear at
$\sim29$ K in high purity Bi$_2$Te$_3$ crystals \cite{Kittel1996}.
It has been demonstrated that the phonon-drag thermopower from a
two-dimensional ($2$D) conducting layer on a three-dimensional
($3$D) crystal can display giant quantum oscillations due to the
phonon intra- and inter-LL scattering in the presence of a strong
magnetic field. In such a $3$D system, surface electrons are
dragged by non-equilibrium $3$D phonons of the whole specimen,
while in a purely $2$D system such as graphene, electrons of a
wavevector $k$ can only interact with $2$D phonons of a wavevector
$q \leq 2k$. In addition, the bulk thermopower is considerably
suppressed due to the existence of two types of bulk carriers with
opposite signs \cite{Qu2010}. Therefore, we expect that the
magneto-thermopower of SSs is orders of magnitude larger than that
of bulk in the high field limit. The thermopower measurement thus
provides a powerful tool to elucidate the nature of the
topological SSs that is difficult to be probed by the conductance
measurement.

\newpage
\begin{figure}
\vspace{0pt} \hspace{-10pt}
\includegraphics[width=3.5in]{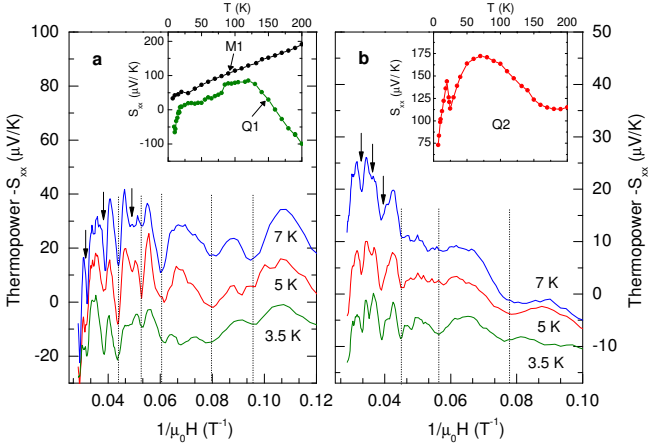}
\vspace{-10pt}\caption{Magneto-thermopower in high magnetic
fields. (a) The thermopower response $-S_{xx}$ versus $1/H$ in
sample Q1 at temperatures $T = 3.5$, $5$, and $7$ K. (b)$-S_{xx}$
versus $1/H$ in sample Q2. The insets show the $T$ dependence of
the thermopower profiles in samples Q1 and M1 in (\textbf{a}) and
in sample Q2 in (\textbf{b}). The arrows mark the sub-integer dips
resolved in $1/H>0.06$ T$^{-1}$ . The dashed lines indicate the
oscillating minima in the low field
regime.}\label{fig:S}\vspace{-15pt}
\end{figure}

Figure 1, a and b, show the thermopower response $-S_{xx}$ versus
the inverse magnetic field $1/H$ in samples Q1 and Q2,
respectively. Large LL oscillations begin to emerge at $H>8$ T and
their amplitude becomes smaller as $T$ decreases from $7$ to $3.5$
K. A prominent feature of these oscillations is that at $1/H
<0.06$ T$^{-1}$ sharp dips (black arrows) appear, with an
aperiodic spacing smaller than the oscillating structure in the low field
limit.

It is illuminating to compare the oscillations in $-S_{xx}$ and
the Shubnikov-de Hass (SdH) effect in the conductance tensor
$G_{xx}$, which was confirmed to arise from the 2D SSs in the
previous study \cite{Qu2010}. As shown in Fig. 2, a and b, the
extrema in $-\Delta S_{xx}$ coincide with the extrema in $\Delta
G_{xx}$. This occurs because both $-S_{xx}$ ($S_{xx}<0$ for
electron-like carriers) and $\sigma_{xx}$ peak when the Fermi
level ($E_\mathrm{F}$) aligns with each LL, whereas vanish when
$E_\mathrm{F}$ lies between LLs. Furthermore, we observe
pronounced LL splitting near $1/H=0.061$, 0.102, and 0.142
T$^{-1}$ (gray dashed lines in Fig. 2b). This splitting indicates
that the degeneracy is lifted between top ($+$) and bottom ($-$)
surfaces. Similar effect has been seen in strained HgTe 3D TIs
\cite{Brune2011}. Here, a weak Te composition gradient in
Bi$_2$Te$_3$ breaks the inversion symmetry and generates displaced
Dirac points (see the inset of Fig. 2a). By cleaving the crystal
into bulk samples with a thickness $t=20\sim100$ $\mu$m, we obtain
slightly different surface carrier densities, which then leads to
two sets of Landau filling factor $\nu$ in one piece of sample
(Fig. 2c). Hence, we can pinpoint the top and bottom surface index
fields $B_{\nu^+}$ and $B_{\nu^-}$ from the periodic spacing of
strong (black dashed lines) and weak (grey dashed lines) minima in
$\Delta G_{xx}$ for sample Q2. Similar results were observed in
sample Q1, with its bottom surface cleaved adjacent to the top
surface of sample Q2. As shown below, the extracted bottom surface
carrier concentration in Q1 is almost the same as the top surface
carrier concentration in Q2. The Fermi energy levels relative to
the Dirac point in samples Q1 and Q2 are sketched in Fig. 2d.

In terms of the carrier concentration $n_e$ on one surface, $B_{\nu}$ is
related to $\nu$ by
\begin{equation}
B_{\nu}=\frac{n_e\phi_0}{(\nu-\gamma)}
\label{eqn:Bv}\end{equation} where $\phi_0=h/e$ is the magnetic
flux quanta, $h$ Planck's constant, $e$ the charge of electron,
and $\gamma$ the filling factor shift. A shift with $\gamma=0$
corresponds to a conventional spectrum, whereas a deviation from
the zero-shift with $\gamma=1/2$ implies a Dirac spectrum. The
$1/2$ arises from the $n=0$ LL at the Dirac point.  In the
following, we label the filling factors as $\nu^s$, where $s=\pm$
indexes the top and bottom SSs. With the $B_{\nu^s}^{-1}$
identified in both $-S_{xx}$ and $\Delta G_{xx}$, we plot them
against integers (triangles and circles in Fig. 3, a and b). The
slopes of the linear-fit to the data yield the carrier
concentration $n_e=9.27$ ($7.61$)$\times10^{11}$ and $7.37$
($5.92$)$\times10^{11}$ cm$^{-2}$, with the Fermi wavevector
$k_\mathrm{F}=0.034$ $(0.031)$ and $0.030$ $(0.027)$
$\mathrm{\AA}^{-1}$, for the top (bottom) SSs in samples Q1 and
Q2, respectively. The linear-fit intercepts the $\nu$ axis at
$\gamma=0.45\pm0.02$ in Q1 and $0.67\pm0.02$ in Q2, consistent
with a Dirac dispersion. Hence, we are again convinced that the
$2$D Dirac states give rise to the LL indexing shown in Fig. 2.
Furthermore, the weak-field Hall anomaly provides an independent
measurement of the average surface wavevector. The value of
$k_\mathrm{F}$ derived from surface Hall conductance is in
reasonable agreement with the quantum oscillation analysis.

\begin{figure}
\vspace{00pt} \hspace{0pt}
\includegraphics[width=3.5in]{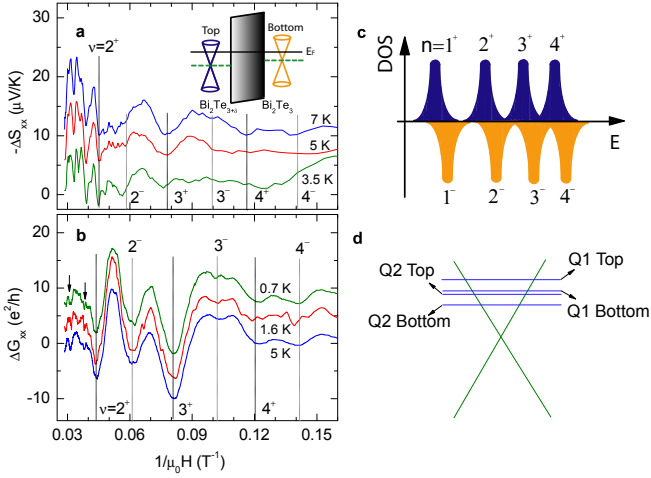}
\vspace{-10pt}\caption{Quantum oscillations from integer Landau
level. (a) The oscillating component of thermopower $-\Delta
S_{xx}$, obtained by subtracting a smooth background from
$-S_{xx}$ in sample Q2. The inset sketches the Dirac cone position
on both surfaces at the existence of a slightly Te composition
gradient in bulk samples. The Fermi energy $E_\mathrm{F}$ is fixed
across the bulk sample. (b) Longitudinal conductivity $\Delta
G_{xx}$ with a smooth background subtracted is plotted against
$1/H$ at selected temperatures $T=0.7$, $1.6$, and $5$ K in the
same sample. Dashed vertical lines mark the minima in $-\Delta
S_{xx}$ and $\Delta G_{xx}$, which are consistent with integer
$\nu$ of top ($+$) and bottom ($-$) Fermi surface. The variation
of Dirac point positions in two surfaces with a constant Fermi
level $E_\mathrm{F}$ is shown as an inset in (a). (c) Sketch
showing how the composition gradient induces the top and bottom
Landau level displacement. (d) Illustration of the energy
difference between the Fermi energy level $E_\mathrm{F}$ (short
lines) and the Dirac point of top or bottom surface states in
samples Q1 and Q2.}\label{fig:Training} \vspace{-15pt}
\end{figure}

We then illustrate the fine structure of $-S_{xx}$ in the range of
$1<\nu<2$ for samples Q1 and Q2 (Fig. 3, c and d). We observed
narrow, reproducible minima at the fields $B_{\nu^s}$ (dashed
lines) with $\nu^s=5/3^{\pm}$. In addition to the minima at
$\nu^s=5/3^{\pm}$, a valley-like structure at $\nu^s=9/5^{\pm}$ is
discernable in Q1, and the dip at $\nu^s=9/5^+$ becomes prominent
in Q2 as the temperature decreases. As seen in Fig. 3, a and b,
the fractional Landau fillings lie on a straight line with the
integer ones. To examine the oscillating profiles of the most
pronounced sub-integer structures, we plot the $-S_{xx}$ versus
the filling factor calculated as $\nu=n_e\phi_0 B^{-1}+\gamma$
(Fig. 4, a and b). The $-S_{xx}(\nu)$ traces obtained at various
$n_e$ are almost overlapped, and their minima are all located
around $\nu=1.67\pm0.02$ (traces are displaced for clarity). Both
features strongly suggest that the observed high-field structures
are associated with the FQH states at $n=1$ LL. Based on the
theory of theomopower in the QHE regime, a minimum in
$\sigma_{xx}$ should accompany a minimum in thermopower. As shown
in Fig. 2b (black arrows), two sub-integer dips are clearly
resolved in $\Delta \sigma_{xx}$ with $\nu^s$ close to $\pm 5/3$.
Noting that for conventional 2D electron gas in the FQH regime,
$\rho_{xy}$ should be quantized with ($\nu e^2/h$)$^{-1}$ at
filling factor $\nu$. Unfortunately, the present of the bulk
conduction channel does not allow us to measure exact Hall
magnitude. Even in most resistive TIs, Hall quantization is
obscured by dominate bulk contribution and no quantized Hall
plateaus have so far been observed.

In spite of the limited data set, we may roughly estimate the
lower bound of the gap energy of the $5/3$ state ($\Delta_{5/3}$)
from the $T$ dependence of the thermopower and resistivity minima,
both of which scale as $e^{-\Delta/2T}$. As shown in Fig. 2b and
Fig. 3, c and d,  the $5/3^+$ state persists until $5$ K,
indicating $\Delta_{5/3} > 5$ K. This value is more than an order
of magnitude larger than the corresponding gap ($\Delta_{8/3}$) in
the GaAs system with a much higher mobility \cite{Pan2008}, but
comparable to the gap in graphene with a similar mobility
\cite{Dean2011}.  This is not surprising because the $n=1$ LL in
the topological SSs is a mixture of the $n=0$ and 1 LLs in
non-relativistic systems. It makes the FQH states in the $n=1$ LL
in the Dirac system more robust than those in the GaAs system
\cite{Apalkov2006,Goerbig2006, Yang2006,Toke2006, DaSilva2011,
Bolotin2009, Du2009}. Compared with graphene, the surface states
in Bi$_2$Te$_3$ has only one Dirac valley with no spin degeneracy,
analogous to a completely four-fold degeneracy lifted graphene
system where the FQH states do not mix between spin- and
valley-bands.  Moreover, the helical spin texture of topological
SSs and the presence of conductive bulk states may lead to
enhanced effective Coulomb interactions, rendering the FQH states
even more robust in TIs \cite{DaSilva2011}.


\begin{figure}
\vspace{00pt} \hspace{-20pt}
\includegraphics[width=3.5in]{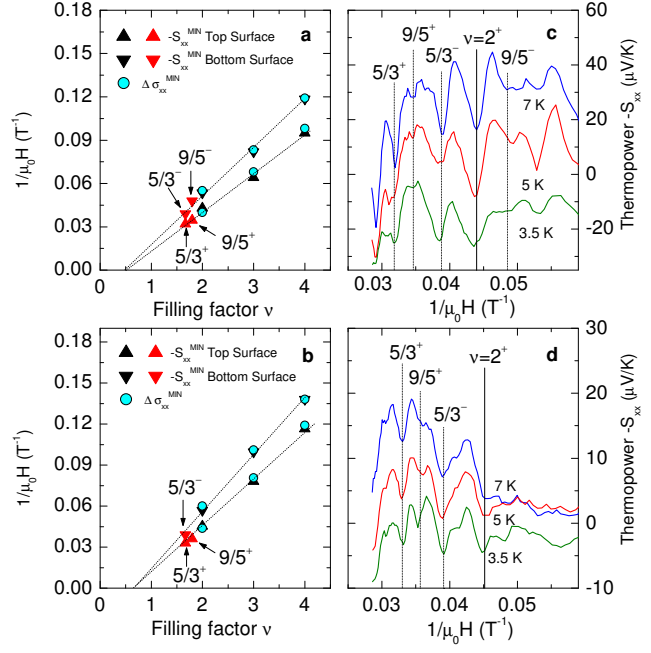}
\vspace{-10pt}\caption{Identification of fractional quantum Hall
states. The left panel shows the $1/H$ positions of the $-S_{xx}$
and $\Delta G_{xx}$ minima versus the filling factor $\nu$ in
sample Q1 (a) and sample Q2 (b). The black and red triangles
represent the integer and fractional fillings, respectively. The
right panel shows the fine structure of thermopower response
$-S_{xx}$ versus $1/H$ for $1/H<0.06$ T$^{-1}$ in sample Q1 (c)
and sample Q2 (d). Identified minima correspond to the
fractional-filling factors illustrated as the vertical dashed
lines. Note that the minima at $\nu=5/3^+$ and $5/3^-$ are much
stronger than those at $\nu=9/5^+$ and
$9/5^-$.}\label{fig:Training} \vspace{-15pt}
\end{figure}

We next extract the surface thermopower $S^s_{xx}$ from the
observed thermopower response. The measured thermopower tensor
$S_{ij}$ can be expressed as the sum,
\begin{equation}
S_{ij}=\sum_{k=x,y}\rho_{ik}(\alpha^b_{kj}+\frac{\alpha^s_{kj}}{t})
\label{eqn:S_tot}
\end{equation}
where $\rho_{ij}$ is the total resistivity tensor,
$\alpha^l_{ij}=\sum_{k=x,y}\sigma^l_{ik}S^l_{kj}$ with $l=b$ or
$s$, the bulk or surface thermoelectric conductivity tensor, and
$\sigma^l_{ij}$ the bulk or surface conductivity tensor. Since
$\rho_{xx}\gg\rho_{yx}$, $\sigma^b_{xx} \gg \sigma^b_{xy}$,
$\sigma^s_{xx} \gg \sigma^s_{xy}$, and $\sigma_{xx} \gg
\sigma^s_{xx}$ for nonmetallic Bi$_2$Te$_3$ in the high-field
regime, $S_{xx}$ can be approximated as
\begin{equation}
S_{xx}=S^b_{xx}+\frac{1}{t}\rho_{xx}G^s_{xx}S^s_{xx}
\label{eqn:S_xx}
\end{equation}
where $S^b_{xx}$ is the bulk thermopower and $G^s_{xx}$ the
surface conductance. The bulk thermopower only gives rise to a
featureless background. The $\rho_{xx}G^s_{xx}/t$ term can be
obtained from the resistivity measurements. We find that the
maximum magnitude of $\rho_{xx}G^s_{xx}/t \sim 0.027$ and $0.01$
in Q1 and Q2, respectively. From Eq. (\ref{eqn:S_xx}), we can
extract the $-S^s_{xx}$ versus $H$ in Q1 and Q2 (Fig. 4, c and d,
red curves). The peak magnitude of $-S^s_{xx}$ is in the range of
$0.5-2.0$ mV K$^{-1}$, which is more than an order of magnitude
higher than that of the bulk $\sim 30$ $\mu$V K$^{-1}$ at $5$ K.
Unlike conventional $2$D systems where the thermopower magnitude
roughly displays a linear field dependence \cite{Fletcher1986},
the surface thermopower at higher order LLs such as $n=4$ is
comparable or even greater than that of lower LLs ($n=3$). This
giant oscillating magnitude and the specific field profile of the
surface thermopower can be understood within the scenario of the
$2$D Dirac electron and $3$D phonon interaction.

Because of the relativistic dispersion of topological surface
states, the wave function $\Psi_{n}$ of a Dirac electron in the
$n$th LL is the superposition of the $n$th and $(n-1)$th LL wave
functions of a non-relativistic electron. The mixture nature of
the wave function significantly modifies the electron-phonon
matrix element in the $n\geq 1$ LLs, leading to a thermopower
profile different from an ordinary $2$D system.

\begin{figure}
\vspace{00pt}\hspace{0pt}
\includegraphics[width=3.25in]{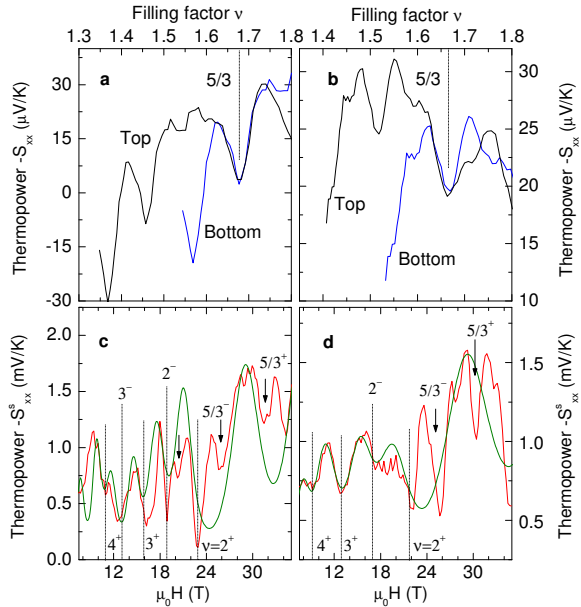}
\vspace{-10pt}\caption{Characteristics of the $\nu=5/3$ state and
the numerical simulation of the surface thermopower. (a) $-S_{xx}$
as a function of the filling factor $\nu=n_e\phi_0 B^{-1}+\gamma$
in sample Q1 at $7$ K. (b) $-S_{xx}$ versus $\nu$ in sample Q2 at
$7$ K. In (a) and (b) curves are vertically offset for clarity.
(c), (d) The surface thermopower $-S^s_{xx}$ obtained by removing
a smooth background contributed from $S^b_{xx}$ in sample Q1 (c)
and sample Q2 (d) at $5$ K. The green curve is the fit to
$-S^s_{xx}$ based on the Dirac fermion and 3D phonon interaction
model. The $H$ positions at the integer-filling factors are marked
by dotted lines. The fractional structures resolved at
$\nu^s<2^{-}$ are indicated by arrows.}\label{fig:Sxx}
\vspace{-15pt}
\end{figure}

Using a general model given in ref. 29 and the wave function for
topological SSs \cite{Liu2010}, we numerically simulate the
thermopower induced by the integer Landau quantization in Q1 and
Q2, with the phonon mean free path treated as a fitting parameter
(Fig. 4, c and d, green curves). We include the average LL
broadening width $\Gamma=3.5$ and $7$ meV for Q1 and Q2,
respectively. The simple electron-phonon interaction model (Eq.
S5) does not capture the fractional features, as FQHE is not
included in the model. However, it reproduces the index field
position and the oscillation magnitude from the integer Landau
quantization. This suggests that the observed giant integer Landau
oscillations can be explained by 2D Dirac fermion and 3D phonon
interaction. A more comprehensive FQHE framework is needed to
model the fractional-filling states in thermopower response.

By performing thermopower measurements, we have resolved
fractional Landau quantization of SSs at $\nu=5/3$ and $9/5$. The
observed gap energy at the $5/3$ state is ten times larger than
that of the non-relativistic electron systems. The demonstration
of the FQH states in the topological surface bands opens the door
to future studies of fractional quantum Hall effect physics in the
topological insulator, which is expected to display strong
correlation effects between chiral Dirac fermions.

The authors would like to thank N. P. Ong, F. D. Haldane, L. Fu,
and C.-X. Liu for helpful discussion.



\end{document}